\documentclass{jpsj2}
\title{Kondo engineering : from single Kondo impurity to the Kondo lattice
}

\author{J. Flouquet, A. Barla, R. Boursier, J. Derr, and G. Knebel}

\inst{CEA-SPSMS, 17 rue des Martyrs, 38054 Grenoble Cedex 9, France}

\abst{In the first step, experiments on a single cerium or ytterbium Kondo impurity reveal the
importance of the Kondo temperature by comparison to other type of couplings like the hyperfine
interaction, the crystal field and the intersite coupling. The extension to a lattice is discussed.
Emphasis is given on the fact that the occupation number $n_f$ of the trivalent configuration may
be the implicit key variable even for the Kondo lattice. Three $(P, H, T)$ phase diagrams are
discussed: CeRu$_2$Si$_2$, CeRhIn$_5$ and SmS. }

\kword{Kondo effect, Kondo lattice, spin fluctuation, magnetism, superconductivity, pseudogap}

\begin{document}
\maketitle

In this article we will use the Kondo idea of strong local enhancement of the effective mass $m^*$
of the quasiparticle to understand the new features which appear in a Kondo lattice i.e for a
regular array of Kondo paramagnetic centers. Thus an attempt will be given to classify the main
parameters of a Kondo impurity and then to schematize the problem by simple relations in order to
advance in the experiments on this complex quantum matter which is the Kondo lattice. For
simplicity, we restrict the discussion mainly to Ce Kondo centers.

\section{Kondo impurity : hierarchy between $T_K$ and other couplings}
The new insight into the Kondo effect is that due to the strong coupling with the Fermi sea below
some Kondo temperature ($T_K$) the entropy of a single paramagnetic impurity can collaps at $T \to
0$ K without further coupling with the other impurities.

 One of the important message given by the study of a single Kondo impurity is that the Kondo
energy $k_B T_K$ is the key energy in the comparison of the electronic coupling given by the Fermi
sea with other sources of interactions like the electronuclear hyperfine coupling $A$
\cite{Flouquet1978} or the crystal field effect $C_{CF}$ \cite{Cornut1972}. If $k_B T_K \leq A$ or
$C_{CF}$, the angular momentum $J$ of the Kondo paramagnetic center will feel these inner
structures, and only after the selection of a reduced  effective spin, the Kondo effect will act and
governs the low temperature properties.

A nice case is the hyperfine coupling between the electronic effective spin $J$ and its nuclear
spin $I$ submitted to the combined effect of a magnetic field $(H)$ and the Kondo exchange term
$(\Gamma)$ between $J$ and $s$ the spin of the conduction electron according to the hamiltonian
\begin{equation*}
H = A\vec I \cdot \vec J + g_J \mu_B \vec H \cdot \vec J + g_n \mu_n \vec H \cdot \vec I -2 \Gamma
\vec J \cdot \vec s
\end{equation*}

\begin{figure}[h]
    \centering
    \scalebox{0.5}{\includegraphics{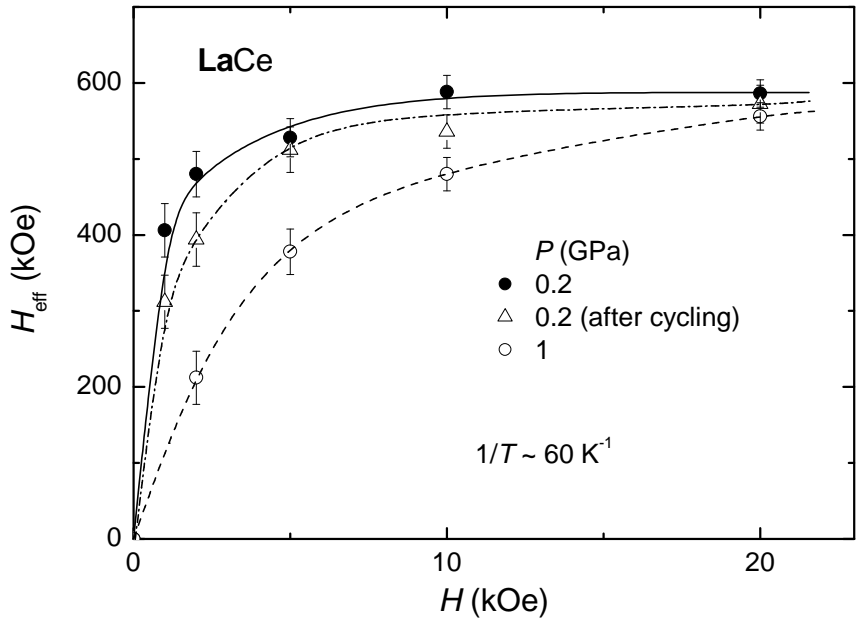}}
    \caption{Nuclear orientation data for \textbf{La}Ce at 0.2 and 1 GPa. The solid and dash dotted curves were
    taken at 0.2 GPa before and after cycling to 1 GPa respectively (after Benoit \textit{et al.}).\cite{Benoit1974}}
    \label{figure1}
\end{figure}

This situation was extensively studied by the nuclear orientation method on highly diluted
radioactive nuclei. So far $k_B T_K  < A$, in a first step $I$ and $J$ are coupled exactly like in
atomic physics. Then the Kondo effect will occur on the selected fundamental state $F = I - J$ or
$I + J$ depending the sign of $A$. When $k_B T_K \geq A$, the behavior of electronic spin will be
first governed by the Kondo effect leading in magnetic field to an average value   $\langle J_z
\rangle$ and thus an effective hyperfine field $H_{eff}$ on the nuclei proportional to the
magnetization. This idea is clearly demonstrated in nuclear orientation experiments realized
notably on \textbf{Au}Ce or \textbf{La}Ce alloys \cite{Flouquet1971}. At zero pressure $(P)$, it was
found for \textbf{Au}$^{137}$Ce that the electronic spin is mainly coupled to the nuclear spin without
any sensibility to any Kondo mechanism (i.e $T_K$ may be lower than mK) why already in \textbf{La}Ce a
large reduction of $H_{eff}$ is observed by comparison to a free electronuclear motion $(k_B T_K
\sim A$). Furthermore, here the measurements under pressure (see figure \ref{figure1})
\cite{Benoit1974} point out an increase of $T_K$ by a factor 5 from 120 mK at 0.2 GPa to 600 mK at
1 GPa. The high degree of localisation of the 4$f$ electron on the 4$f$ shell (occupation number
$n_f$) is also nicely proved by the weak variation of the saturated value.

\begin{figure}[h]
    \centering
	\scalebox{0.5}{\includegraphics{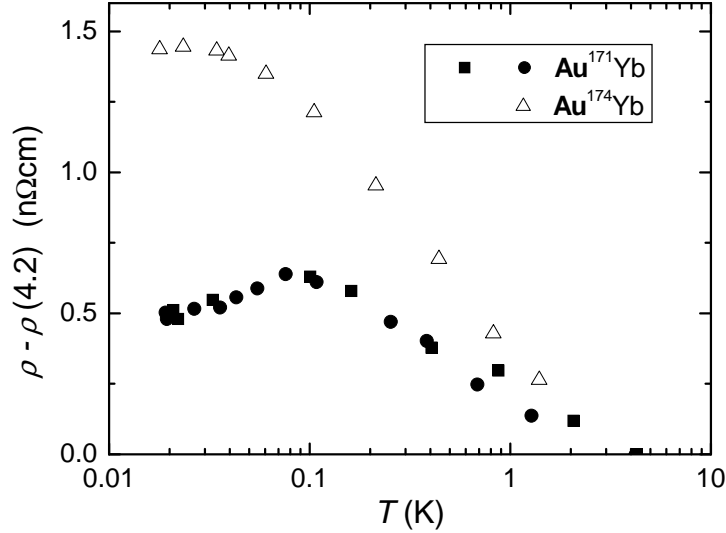}}
    \caption{The square and triangle points describe the \textbf{Au} $^{171}$Yb resistivity of a 415 ppm
    alloy, the circle the \textbf{Au}$^{174}$Yb resistivity of a 396 ppm alloy. The residual logarithmic
     slope of \textbf{Au}$^{171}$Yb reflects only the presence of parasitic 5 ppm impurities of ion. The
      important feature is for \textbf{Au}$^{171}$Yb, the maximum in the Kondo scattering coincides with
       the temperature where the singlet electron nuclear state becomes isolated from the triplet excited
       level (after Hebral \textit{et al.}).\cite{Hebral1977}}
    \label{figure2}
\end{figure}

An illustration of the interplay between the Kondo effect and the hyperfine coupling was given by the study
of the resistivity of \textbf{Au}Yb realized with different isotopes and thus different hyperfine
couplings \cite{Hebral1977}. Hyperfine measurements either by nuclear orientation
\cite{Benoit1974a} or M\"{o}ssbauer effect \cite{Gonzalez1973} push the limit of $T_K$ below the
millikelvin range. The element Yb has a number of stable radio-isotopes ; $^{171}$Yb and $^{173}$
have nuclear moments with opposite spin. By choosing the $^{171}$Yb isotope, with spin $I = 1/2$,
the positive sign of $A$ leads to a zero angular momentum $F = I - J = 0$. Resistivity measurements
on figure \ref{figure2} show the breakdown of the Kondo scattering for Au$^{171}$ Yb by comparison
to the case of \textbf{Au}$^{174}$Yb (even nuclei with no nuclear moment).

A highly discussed problem is the interplay between $k_B T_K$ and $C_{CF}$. Experiments notably on
\textbf{Al}$_2$\textbf{La}Ce (see \cite{Cornut1972,Steglich1977} have shown also that provides $k_B T_K
\leq C_{CF}$, the Kondo effect at low temperature is realized on the doublet ground state
$\Gamma_7$ for this cubic material. Under pressure, as we will see later, the Kondo temperature
will increase strongly while the crystal field splitting has a smooth pressure variation.
The recovery of the full degeneracy under pressure at a pressure $P_V$ is an important issue in the
change of a single Kondo behavior but also in the properties of the Kondo lattice.

When the impurity cannot be considered as isolated among the Fermi sea, one must discuss the
interplay between the Kondo energy and the intersite coupling $E_{ij}$. For an effective spin equal
to $1/2$ i.e for a high degree of the 4$f$ localisation $(n_f \sim 1)$, $E_{ij}$ can be mediated
by the well known Ruderman, Kittel, Kasuya, Yoshida interaction (RKKY) carried out by the light
electron. As a function of $\Gamma$ and $N(E_F)$ the density of states of the light electron, $k_B
T_K$ and $E_{ij}$ can be written \cite{Hewson1992} as
\begin{align}
E_{ij} &= \Gamma^2 N(E_F) \nonumber \\ k_B T_K &\sim \frac{1}{N(E_F)} \exp\left(-\frac{1}{\Gamma N(E_F)}\right)
\label{eqn1}
\end{align}
with $\Gamma N(E_F) = \Delta / E_0$ and $k_B T_F = N(E_F)^{-1}$. $\Delta$ and $E_0$ correspond
respectively to the line width and position of the virtual 4$f$ bound state. These two terms fix
$n_f$. When $k_B T_K \geq E_{ij}$ at a critical pressure $P_C$, the long range magnetism will
collapse according to the popular Doniach scheme \cite{Doniach1977}. However, the effects in the
Kondo lattice may be more subtle, when $k_B T_K$ crosses $C_{CF}$ at $P_V$, $E_{ij}$will drop by one
order of magnitude \cite{Ramakrishnan1981}. Thus $P_C$ can never exceed $P_V$. Of course, now, the
electronic conduction bath is not infinite by respect to the number of cerium sites.

\section{From the impurity to the lattice}
Three different views can be taken for the collapse of long range magnetism : (1) the previous
Doniach scheme \cite {Doniach1977}, (2) the generalisation of the spin fluctuation theory successfully
applied for 3d itinerant magnetism \cite{Moriya1995}, (3) a push pull mechanism between magnetic and
electronic instability strongly related with a change in the occupation number. In the case (1), the
Fermi surface is basically assumed to be that given under the assumption of a localized 4f electron.
In case (2), the 4f electron is considered to be already itinerant at a pressure $P_{KL} \leq
P_C$. However, in case (3), the quantum effects are furtive with the interferences between spin dynamics, charge
motions and their localization.

If the collapse of long range magnetism is via a second order phase transition, one may expect a
divergence of the magnetic coherence length $\xi_m$ at $P_C$ coupled with a collapse of a
characteristic temperature. In spin fluctuation theory, the corresponding temperature $T_{sf}$
vanishes as $(P - P_C)^\alpha$ with $\alpha = 1$ or $3/2$ for antiferromagnetism or ferromagnetism.\cite{Moriya1995}
If the hidden parameter is the occupation number, one may expect a tiny discontinuity in volume and
thus a first order transition with the consequence of finite value of $\xi_m$ and related low Kondo
lattice energy $T_{KL}$.

\begin{figure}[h]
    \centering
	\scalebox{0.5}{\includegraphics{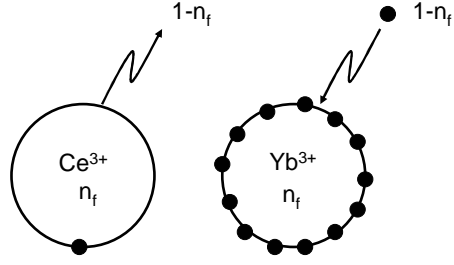}}
    \caption{It is the tiny departure from $n_f = 1$ by the release of $(1 - n)$ 4$f$ electrons for
    the trivalent Ce impurity or the absorption of $1 - n_f$ electrons for Yb ion which leads to the Kondo effect.}
    \label{figure3}
\end{figure}

Figure \ref{figure3} mimics the situation of Ce$^{3+}$ and Yb$^{3+}$ which corresponds to the
trivalent configuration 4$f^1$ and 4$f^{13}$. The Kondo phenomena is linked with the release of $1
- n_f$ electron from Ce$^{3+}$ or the addition of $1 - n_f$ electron to Yb$^{3+}$ according to the
scheme:
\begin{align*}
\rm{Ce}^{3+} &\leftrightarrows \rm{Ce}^{4+} + 5d \\ \rm{Yb}^{2+} &\leftrightarrows \rm{Yb}^{3+} + 5d
\end{align*}

For a single impurity, the expression of $T_K$ as a function of the occupation number is given by \cite{Hewson1992}
\begin{equation}
\label{eqn2}
k_B T_K \sim \frac{1-n_f}{n_f} \Delta
\end{equation}
of course $T_K$ depends as $n_f$ of the value of $E_0$ and $\Delta$ (the degeneracy factor $N_f$ has been dropped). In a lattice, there will be a
feedback between the Kondo effect on a given site and the Fermi sea. That pushes to propose for the lattice a simple form for $T_{KL}^{3+}$ equal to :
\[
T_{KL}^{3+} = (1 - n_f)T_F
\]
with $T_F \sim D(n_F)$ a typical electronic bandwidth. $T_{KL}^{3+}$ is  attributed to the 3$+$
configuration (Ce$^{3+}$ or Yb$^{3+}$) during its lifetime. (The missing factor $n_f$ in formula \ref{eqn2} will be recovered as the average susceptibility will vary as $n_f \chi_{3+} \sim T_{KL}^{-1}n_f$.)

Looking only to the low carrier doping
(i.e neglecting the Brillouin zone). The corresponding relation for $T_{KL}$ for Ce$^{3+}$ and
Yb$^{3+}$ configuration will be:
\begin{align}
k_B T_{KL} ({\rm Ce}^{3+}) &= (1 - n_f)^{5/3}  D_0 \\ 
k_B T_{KL} ({\rm Yb}^{3+}) &= (1 - n_f)
n_f^{2/3} D_0
\label{eqn3}
\end{align}
where $D_0$ is the  5$d$ bandwidth and independent of $n_f$. A sound discussion on the relation
between $T_{KL}$, $n_f$ and $n_c$ the extra conduction number in the lattice can be found in
Trees \textit{et al.},\cite{Trees1995} and Ikeda and Miyake \textit{et al.}\cite{Ideka1997}. In this crude model, the striking point is
that the dependence $T_{KL}$ for the Ce$^{3+}$ configuration will increase continuously as $n_f$
decreases. For the Yb$^{3+}$ one, as the original free 5$d$ electron becomes absorbed in the
4$f^{14}$ configuration, $T_{KL}$ will go through a maximum. Thus slow relaxing Yb$^{3+}$ (but also
Sm$^{3+}$) moments characteristic of the trivalent configuration can survive even for a rather weak
occupation number ($n_f \sim 2.7$). We want to point out, that one can expect drastic differences between
Ce and Yb (or Sm) compounds in the magnetic properties near $P_C$. Furthermore, the underlining
idea is that the effective mass $m^*$ will be finite even at $P_C$ and is basically determined by $T_K$.
The non-divergence of $m^*$ appears clearly in the superconducting phase coupled with the collapse of the antiferromagnetic ordering at $P_c$. Rather moderate values of the strong coupling parameter is required for a good fit of the upper critical field.

 but the
associated coherence length $\ell_{LK}$ associated with $T_{KL}$ and $m^*$ can strongly increase near
$P_C$. Furthermore, the 5$d$ electrons may react to the 4$f$ correlations. So $D$ may depend on a
less trivial form on $n_f$. For example, $D(n_f)$ may reach a high value for $n_f \rightarrow 1$ and
a weak one for $n_f \leq 2.8$. An illustration will be given below with the gold phase of SmS where
obviously at small enough pressures $(P<P_\Delta$ see later) the 5$d$ electron prefers to be trapped
around the Sm centers. The relation of the pseudogap shape of $D$ with $n_f$ is a central issue in
strongly correlated electronic systems and will not be discussed here.

The key problem is to derive a characteristic temperature for the lattice. It is amazing that in the limit of low carrier density it was proposed that $T_{KL}$ varies as $n^{1/3}$.\cite{Burdin} With the previous relation 
\ref{eqn3}, this prediction is recovered in the Yb case remaining that $T_{KL}=T_{KL}(\rm{Yb}^{3+})/n_f$. Another suggestion was that $T_{KL}$ goes like $T_K^2$,\cite{Nozieres1998} an expression not so far to the initial $T_K^
{5/3}$ law proposed for the Ce case. An appealing possibility is that as the effective mass increases, its 
coherent circulation over the Kondo loop increases. If $\ell_{KL}$ is proportinal to $m^*$, the heavy particle of mass $m^*$ will travel through the full Kondo loop after a time $\tau_{KL}\sim (m^*)^2$. Of course the great 
novelty is that the situation is different from that found in normal rare earth intermetallic compound where the information is carried by light particles with mass near the bare electronic mass $m_0$ over a small atomic 
distance $a_0$ leading to a fast response $\tau_0 \sim m_0 a_0$. This slow motion over large distances may lead 
to unusual time and space response of the quasiparticle.

\section{Pressure effects}
Discussing the pressure variation of $T_K$ for a single impurity requires to know the relative
pressure variation of $\Delta$ and $E_0$. The expression for $\Delta$ as a function of the hybridization
term $v_{dk}$ between the $f$ and light electron is :
\begin{equation*}
\Delta = \pi v_{dk}^2 N(E_F)
\end{equation*}
The pressure increase of $\Delta$ is related to the increase of the bandwidth of the Fermi sea. It
is wellknown for normal metal that the volume variation of $T_F$ is governed by the increase of the atomic
density leading to a Gr\"{u}neisen parameter $-\frac{\partial \log T_F}{\partial \log V} = \Omega_F
= + 2/3 $. If we assume $v_{dk} \sim N(E_F)^{-1}$ one recover the formula suggested for the Kondo
lattice as $\Delta \sim T_F$. The high sensitivity of the cerium impurity to $P$ can be easily understood
(assuming $E_0$ fixed) by the weakness of $\Delta/T_F \sim 10^{-2}$ in comparison to
$E_0/T_F \sim 0.2$. After equation \ref{eqn1}, with $T_K$ equal to:
\[
T_K= T_F \exp\left(-\frac{E_0}{2\cdot 10^{-2} T_F}\right)
\]
the Gr\"{u}neisen parameter of $T_K$ becomes
\[
\Omega_{T_K} = \Omega_{T_F} \left(1+ \frac{50E_0}{T_F}\right) \sim 10 \Omega_{T_F}.
\]
If $E_0$ is reduced of course there will be an extrasource of an amplification.

Another evaluation of $\Omega_{T_K}$ can be made via the relation \ref{eqn2}. Of course $T_K$
depends on two parameters, $n_f$ and $\Delta$ but assuming $\Delta = T_F$, the main source of
variation of $T_K$ arises through that of $n_f$:
\[
\frac{\partial T_K}{T_K} \sim \frac{1}{(1-n_f)}\frac{\partial n_f}{n_f}
\]
with the crude estimate that $n_f$ varies linearly with the volume according to Vegard's law :
\[
n_f=\frac{V-V_{4+}}{V_{3+}-V_{4+}}
\]
$V_{3+}$ and $V_{4+}$ being the volume attribute to the $3+$ and $4+$ configuration. One gets
\[
\Omega_{T_K} \sim \frac{1}{(1-n_f)}\frac{V_{3+}}{(V_{3+}-V_{4+})}\; {\rm for}\; n_f \to 1;
\]
For $V_{3+} - V_{4+} = 0.5 V_{3+}$ (the $4f$ configuration is in fact never reached for the cerium case)
, the Gr\"{u}neisen parameter of $T_K$ will be $\Omega_{T_K} \sim 2(1-n_f)^{-1}$. For cerium heavy fermion compounds $n_f \stackrel{>}{\sim}0.9$, thus $\Omega_{TK} \stackrel{>}{\sim} 20$.

Basically, $\Omega_{T_K}$ and $m_K^*$, the local effective mass due to the Kondo effect $m_K^* \sim \frac{1}{T_K}$ scale each
other. As pointed out \cite{Jaccard1985}, a constant ratio $\Omega/m_K^*$ will imply a logarithmic
dependence of the interaction responsible for the dressing of the quasiparticle. Recent
developments in high energy spectroscopy of Kondo systems have lead to a direct charge visualization of the
Kondo resonance for cerium intermetallic compounds. However, the extrapolation to the single
impurity parameter is not straightforward (see Malterre et al.\cite{Malterre1996}). Let us stress
here that qualitatively the results obtained on Ce compounds agree with our previous statements. The
situation is far less clear for the the Yb systems. It was already underlined that the differences
reflect a smaller interconfiguration energy in the Yb case than in the Ce one. For Ce, the 4$f^0$
and the 4$f^1$ configurations are separated by 2 eV. For Yb, the separation can be less than 100
meV.\cite{Malterre1996}

In temperature, the experimental Gr\"{u}neisen parameter defined as the ratio of the thermal expansion $\alpha =
-\frac{1}{V}\frac{\partial V}{\partial T}$ by the specific heat $C$ (normalized by the molar volume
$V$ and the compressibility $\kappa$) is a simple powerful tool to test if the free energy $F$
depends only on one parameter ($T^*$, here $T_K$ for a single impurity) as if :
\begin{align*}
F &= T \Phi (T/T^*)\\
 \Omega (T) &= \Omega_0 = constant \nonumber \\
 \Omega (T) &= -\frac{\partial
\log T^*}{\partial \log V} = \frac{\alpha}{C}\frac{V}{\kappa}
\end{align*}

Typical values for Kondo impurity are around 50. The remarks of large Gr\"{u}neisen parameters for
heavy fermion compounds were made in.\cite{Benoit1981}, \cite{Takke1981} The thermal expansion
is huge as it goes then as $(m^*)^2$. If $m^*$ is three order larger than the bare electronic mass,
the corresponding electronic thermal expansion is six times higher than the usual electronic one.
That leads us to suspect that the density fluctuations are crucial in the heavy fermion problems and notably in Cooper pair pairing.

When we will consider the lattice, the study of the ratio $\alpha/C$ in temperature is a sound
cheap way to test if competing interactions occur. The slow recovery of a constant $\Omega (T)$ is
the indication to the "`tenuous"' entrance in a Fermi liquid regime. In spin fluctuation theory,
the effective mass increases at $P_c$ but stays finite while $\Omega^* (P_c)$ diverges
\cite{Moriya1995,Zhu2003}.

\section{CeRu$_2$Si$_2$ : an example of paramagnetic Kondo lattice}
\begin{figure}[h]
    \centering
      \scalebox{0.5}{\includegraphics[clip=,angle=0]{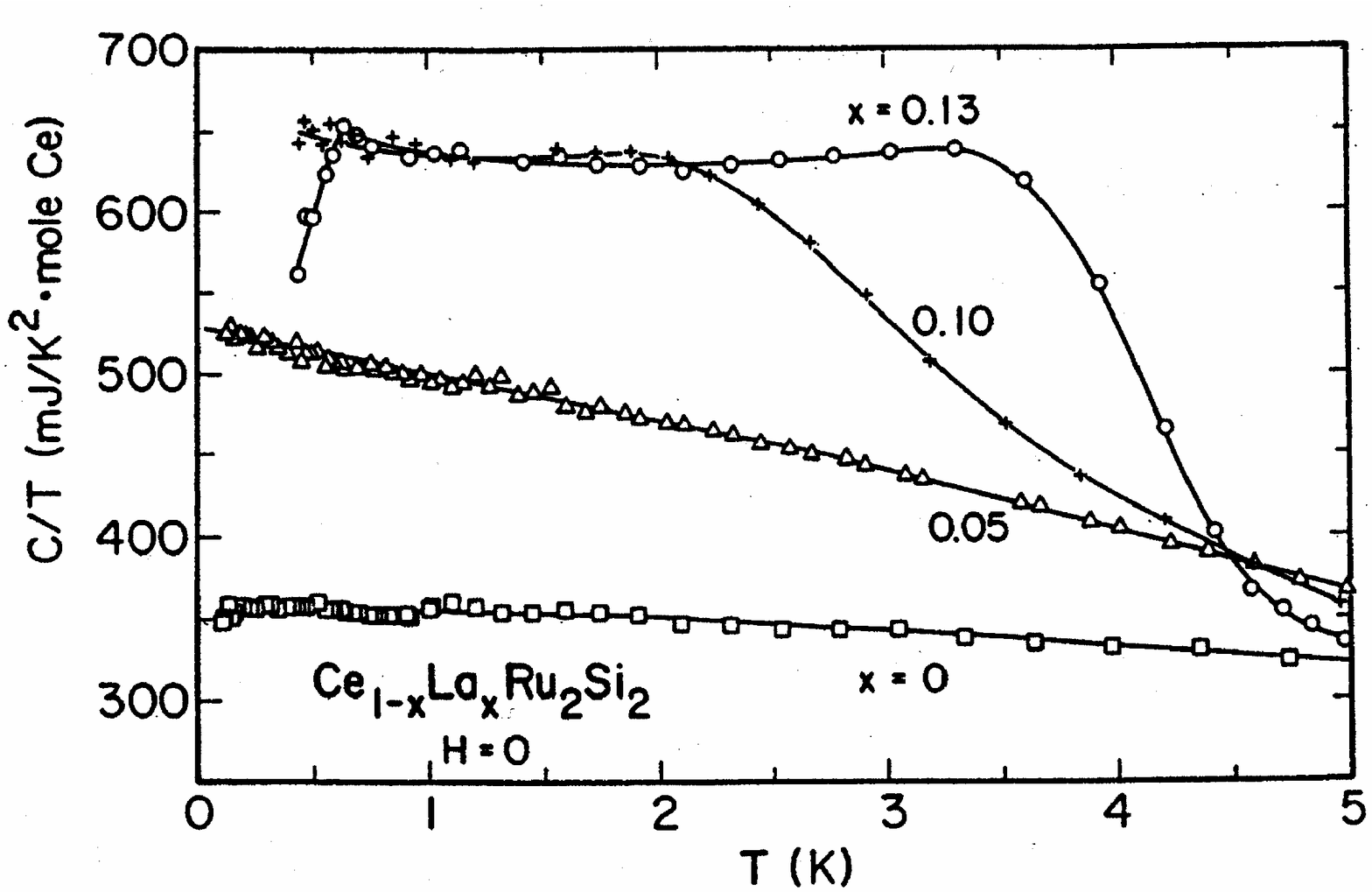}}
    \caption{Specific heat of Ce$_{1-x}$La$_x$Ru$_2$Si$_2$ on the both side of the critical concentration
     $x_c = 0.075$. For $x = 0.13$, antiferromagnetism is observed. For $x = 0.05$,  paramagnetism exists
     down to 0K.}
    \label{figure4}
\end{figure}

The tetragonal compound CeRu$_2$Si$_2$ at zero pressure realized ideal conditions for the
understanding of a high correlated Kondo lattice : $P_C \sim - 0.3$ GPa will be located few kbar
below $P = 0$, furthermore $P_V$ is near 4 GPa \cite{Flouquet2002}. So a large pressure range
exists for a careful study of a spin 1/2 Kondo lattice. The tuning below $P_C$ can be even achieved
by expanding the lattice with lanthanum \cite{Fisher1991} or germanium substitution. In the serie
Ce$_{1-x}$La$_x$Ru$_2$Si$_2$, antiferromagnetism (AF) is only detected for $x \geq 0.075$. The
figure \ref{figure4} represents the specific heat data for  well ordered compounds $x > 0.075$ and
a paramagnetic ground state $x = 0.075$. In the $x =0.05$ paramagnetic case, the slow increase of
$C/T$ on cooling reflects the so called large crossover non Fermi liquid regime. The usual Fermi
liquid dependence of $C/T$ for a nearly antiferromagnet
\begin{equation*}
\frac{C}{T} = \gamma + \beta T^2
\end{equation*}
will only by recovered at very low temperature.
\begin{figure}[h]
    \centering
     \scalebox{0.5}{\includegraphics{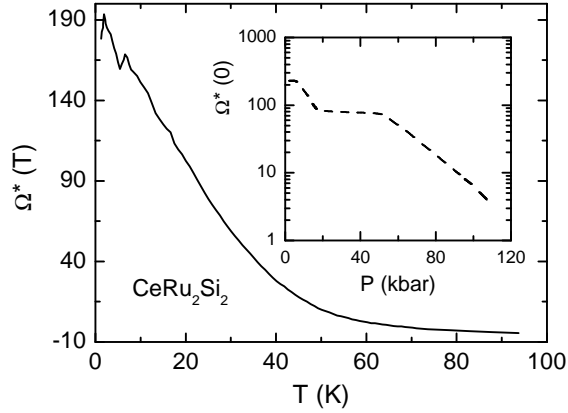}}
    \caption{Temperature variation of the Gr\"{u}neisen parameter measured by the
      normalized ratio of the thermal expansion by the specific heat. In inset, pressure varuation of the Gr\"{u}neisen parameter extrapolated to $T=0$ K.}
    \label{figure5}
\end{figure}
By contrast to the single Kondo effect governed by a single parameter $T_K$, over a large
temperature range $T\lessgtr T_K$, the slow development of the coherence between the spin and
charge motion is pointed out in the temperature evolution of $\Omega^* (T)$ \cite{Lacerda1989}
which reaches a constant only at very low temperature (fig. \ref{figure5}). The complexity in the
continuous process for the formation of the heavy quasiparticle on cooling emerges in the different
macroscopic  experiments. Large differences appear in the characteristic temperature as $T_{-,+}
\sim 60$ K the temperature where the initial magnetoresistance goes from negative to positive,
$T^{M}_{C}\sim 12$ K, $T^{M}_{\chi} = 10$ K, $T^{M}_{\propto} = 8.7$ K the temperature of the
maxima in specific heat, susceptibility and thermal expansion, $T_Q \sim 0.5$ K where the
thermoelectric power $Q$ reaches its $T$ linearity and $T_A = 0.3$K where the Fermi liquid
inelastic $AT^2$ term is observed in resistivity \cite{Flouquet2004}.

The excellent quality of large crystal have allowed two important microscopic experiments.
Inelastic neutron measurements show clearly that antiferromagnetic correlations set in at $T_{-,+}$;
even at $T \to 0$ K, the coherence length is restricted to few atomic distances
\cite{Flouquet2004}. De Haas van Alphen experiments needs the itineracy of the $f$ electrons to be
explained selfconsistently with band calculations at least $P_{KL} \leq P_V$. An effective mass up
to 120 m$_0$ has been detected \cite{Aoki1995,Julian1994}. Applying pressure in CeRu$_2$Si$_2$
leads to a strong decrease of $m*(\Omega^* (P=0) \sim 190)$ \cite{Payer1993}. The derivation of the
pressure variation of the Gr\"{u}neisen parameter shows that $\Omega^*$ reaches another plateau at
$P_V \sim 4$ GPa with $\Omega^*(0) = 80$.

\begin{figure}[h]
    \centering
   \scalebox{0.5}{\includegraphics[clip=,angle=90]{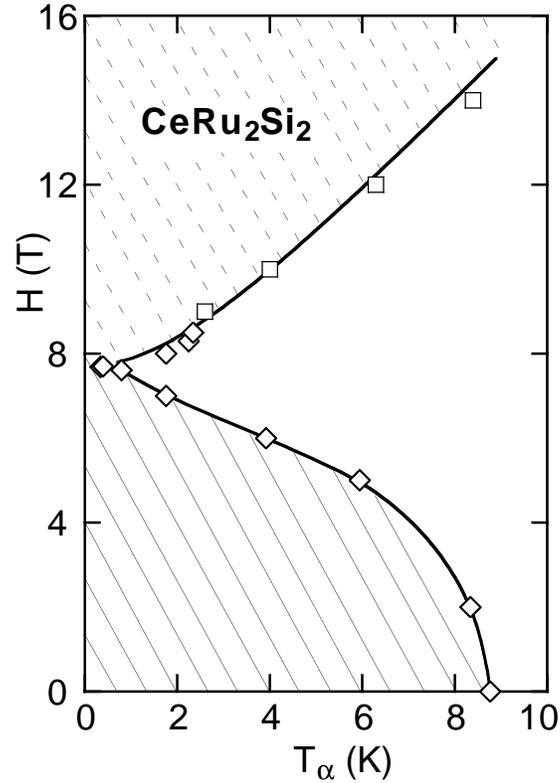}}
    \caption{Field dependence of the temperature of the optimum of the thermal expansion measured
    at different field $\alpha_H (T)$. The full line describes the low field paramagnetic phase dominated
     by AF correlations, the dashed lines the high field phase dominated by a F coupling.}
    \label{figure6}
\end{figure}

An external magnetic field can modify the nature of the magnetic interactions. The metamagnetic
field $H_C$ characteristic of the AF $(H, T)$ phase diagram ends up at a critical point at
$P_C$\cite{Haen1996}. For $P \geq P_C$, only a well defined crossover at $H_M$ will occur between a
low field state dominated by AF correlations $(H \leq H_M)$ and a high field phase $(H \geq H_M)$
dominated by ferromagnetic (F) coupling and the progressing alignment of the cerium magnetization.
That leads to drastic changes in some characteristic temperatures as T$_\propto$ \cite{Fisher1991}.
Figure \ref{figure6} represents the deep minima reached by $T_\alpha$ at $H_M$ : T$_\alpha (H_M)
\sim 0.3$ K i.e roughly the value of $T_A$ at $H$ = 0. The switch from AF to F correlations
\cite{Flouquet2004}, \cite{Raymond1999}, \cite{Sato2004} have been nicely observed in neutron
scattering experiments as well as associated drastic changes of the Fermi surface through $H_M$
\cite{Aoki1995,Julian1994}.

It is worthwhile to underline that $P$ and $H$ are generally different. Under pressure here, the
dominant AF correlations are preserved, but slowly mollified on approaching $P_V$. So the phenomena
at $P_c$ for $H<H_M$ can be ranked as AF quantum critical behavior. In $H$ there is a switch in the nature of
the correlations at $H_M$ which emerges from a critical end field $H_c (P_c)$.
However the change from AF to F interactions is governed by the $H$ shift of the Fermi level inside
the pseudogap structure created in this Kondo lattice \cite{Satoh2001}. Thus the quasi divergence
of $\Omega^* (H,P=0)$ occurs at $H_M$ \cite{Flouquet2004}. If $H_c$ will continuously collapse with
$T_N$ as $P \to P_c$, the magnetic field opens the possibility to tune at low magnetic fields
through $P_c$ \cite{Custers2003}. A rich variety of$(H,P,T)$ puzzles with new figures can be drawn with the interplay of the crossover phase diagram of figure \ref{figure6} and classical phase diagrams for magnetism and supraconductivity

Towards these clear facts, the paramagnetic phase of CeRu$_2$Si$_2$ may be the matter of new
motions (magnetic current) as a tiny magnetism (sublatice magnetization near $10^{-3} \mu_B$) has
been detected around $T \sim 2$K \cite{Amato1993} with an associated signature in the
thermoelectric power \cite{Amato1989}.  An appealing possibility of this furtive magnetism is that
it will be a macroscopic consequence of the circulation of the quasiparticule along their special Kondo
lattice loop. Another open question is the validity of the concept of a second order phase
transition. New experiments on Ce$_{0.925}$La$_{0.075}$Ru$_2$Si$_2$, which is located at $x=x_c$, shows clearly that the characteristic energy does not collapse \cite{Knafo} at $x_c$. The dogma of a second order quantum
critical point may be revisited. The interesting perspective is that the magnetic instability is
associated with a change in the localization of the particle (i.e in a discontinuity of $n_f$).
Indeed for CeIn$_3$ \cite{Knebel2002}, CePd$_2$Si$_2$ \cite{Demuer2002}and CeRh$_2$Si$_2$
\cite{Araki2002}, $P_C$= $P_V$ and first order transitions at $P_c$ are observed\cite{Knebel2002,Araki2002} or strongly suspected\cite{Demuer2002}. Of course, the transition may be weakly first order as large fluctuations are generally observed in the specific heat. It is worthwhile to compare with the first order liquid-solid transition of $^3$He where on the melting curve at $P\sim 34$ bar the volume discontinuity is near 5\%. For heavy fermions, $P_c$ is often around 3 GPa, i.e. three orders of magnitude larger than in $^3$He. Thus a volume discontinuity of $5\times10^{-5}$ will give a comparable mechanical work $(P\Delta V)$ than in $^3$He on its melting curve. Tiny first order transitions may correspond to a $10^{-7}$ or less effect on the volume contraction. 

To go back on the hyperfine coupling, it is worthwhile to point out that for Ce only an even stable
isotope exists. There is no hyperfine coupling. However for natural Yb, four isotopes exist two
even $(I = 0)$ $^{172}$Yb and $^{174}$Yb but two odd, $^{171}$Yb $(I = 1/2)$ and $^{173}$Yb $(I =
5/2)$. If $T_{KL} \leq A$, the full electronuclear dynamic must be considered. Even, an unusual
magnetic ordering may be boosted by the nuclear spin. This may lead to new features for the Yb compounds in comparison to the cerium heavy fermion compounds. Surprisingly for YbRh$_2$Si$_2$ even with a very low N\'{e}el
temperature $T_N \sim 70$ mK and tiny ordered moments (see \cite{Sato2004}), a very nice specific heat anomaly occurs while in Ce Kondo lattice, tiny calorimetric signature is only visible \cite{Flouquet2004}. Recent
experiments on YbRh$_2$Si$_2$ with the $^{174}$Yb isotope shows the persistence of magnetic
ordering.\cite{Glazkov2004} The different behavior between the CeRu$_2$Si$_2$ serie and YbRh$_2$Si$_2$ may be due to
4$f$-5$d$ correlations not included in the usual Kondo approaches.

\section{Magnetism and superconductivity}

In heavy fermion compounds, superconductivity appears related with magnetic and valence
instabilities ($P_C$ and $P_V$). The $P$ and $H$ tunings of the electronic correlations
give unique tools in the interplay between magnetism and superconductivity. Furthermore due to the
weakness of the parameters ($T_{KL} \sim 1$ K, $H_M \sim 10$ T) experiments can be realized with
small scale equipments. In the previous case of CeRu$_2$Si$_2$, no superconductivity has been
detected under pressure. That may emphasize at $P_C$ either the weakness of the pairing mechanism
for Ising spin as Ce in CeRu$_2$Si$_2$ \cite{Flouquet2004} or as observed for CeRh$_2$Si$_2$
\cite{Araki2002} a sharp superconducting domain right at $P_C$. Around $P_V$, new careful pressure
experiments must be realized to test if another superconducting domain may exist as found for
CeCu$_2$Si$_2$ \cite{Holmes2004} and CeCu$_2$Ge$_2$ \cite{Jaccard1992}.

Recently, careful ac calorimetric experiments \cite{Knebel2004} were done on the new discovered 115
CeRhIn$_5$ compound \cite{Thompson2001}. The great experimental interest of this material is that
the optima of their magnetic ordering at $T_N$ and superconductivity at $T_C$ temperatures are
comparable and located in a easy range of temperature ($T \sim 2$ K). As shown in figure
\ref{figure6}, AF disappears via a first order transition. In good agreement with this statement,
AF never succeeds to superconductivity on cooling. Below $P_C$, the opposite occurs but the
superconducting phase is gapless. Our proposal is that specific heterogeneities occurs when $T_N$
starts to decrease deeply. Here bulk superconductivity and antiferromagnetism may be antagonist.

\begin{figure}[h]
    \centering
    \scalebox{0.5}{\includegraphics{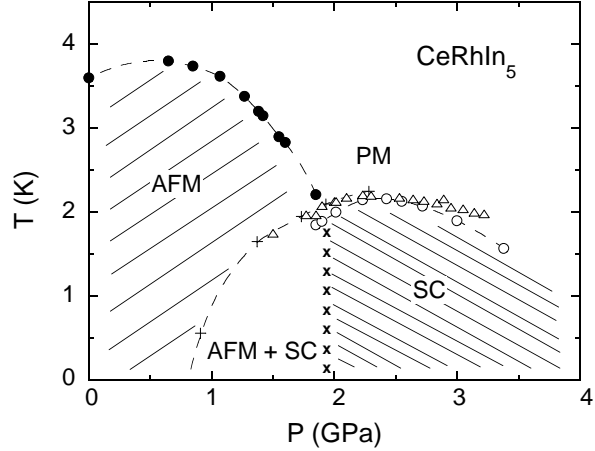}}
    \caption{Phase diagram of CeRhIn$_5$ as determined by specific heat ($\bullet$ and $\circ$)
     and susceptibility ($\triangle$). In addition T$_C$(P) from resistivity measurements after \cite{Llobet2004}.
     ($\times$) marks the first order phase boundary between antiferromagnetism (AF) and superconductivity (SC).}
    \label{figure7}
\end{figure}

\section{Valence, conduction and magnetism : SmS}
One of the fascinating facet of the Kondo lattice illustrated in SmS experiments
\cite{Thompson2001}, \cite{Llobet2004}, \cite{Barla2004}, is that the physical properties may look
that of a specific configuration Sm$^{2+}$ ($J = 0$) or Sm$^{3+}$ ($J = 5/2$) despite the fact the
occupation number $n_f$ if far from unity. New results on SmS demonstrate the curious dressing of
the heavy electrons. SmS is a key matter as the previous equilibrium:
\[
{\rm Sm}^{2+} \leftrightarrows {\rm Sm}^{3+} + 5d
\]
governs the release of a 5$d$ itinerant electron.

In the divalent black (B) phase, Sm$^{2+}$ S$^{--}$ is an insulating non magnetic as $J = 0$ for
the 4$f^6$ Sm$^{2+}$ configuration. In the intermediate valent gold (G) phase, the valence $v$ is
near 2.7 $(v = 2 + n_f)$ right at the first order phase transition at $P_{B-G} \sim 1.2$ GPa ; the
high temperature behavior corresponds to the disordered regime of a Kondo lattice but at low
temperature up to $P_\Delta \sim 2$ GPa the ground state ends up in a non magnetic insulating phase
: $D$ depends on $n_f$. Above $P_\Delta$ for $v \sim 2.8$, SmS is a metal at $T \to 0$ K ; its
metallic behavior is quite analogue to that reported for ordinary heavy fermion compounds.

\begin{figure}[h]
    \centering
   \scalebox{0.6}{\includegraphics{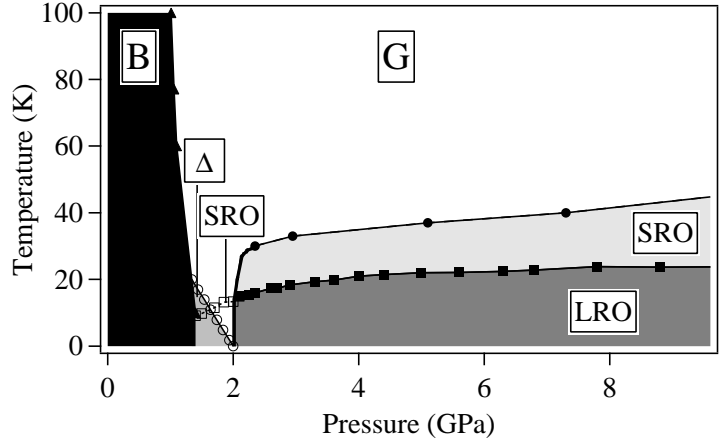}}
    \caption{Phase diagram of SmS in the gold phase. Slow relaxation is observed in NFS experiments
     at $T_{NFS}$, ac calorimetric measurements has located the ordering temperature $T_N$. Short
     range magnetic correlations persists in the low pressure $(P \leq P_\Delta)$ of the gold phase where a gap occurs for the electronic conduction. Above $P_V$, the ground state is AF and metallic. The trivalent state is reached far above 10 GPa \cite{Barla}, \cite{Rohrer1982}.}
    \label{figure8}
\end{figure}

The simultaneous achievement of a macroscopic ac specific heat measurements \cite{Barla2004} and of
a microscopic nuclear forward \cite{Haga2004}, \cite{Barla} scattering experiments (NFS) gives now
a complete view of the interplay between valence, electronic conduction and magnetism. Right at
$P_\Delta$, evidences were found below NFS for the slow spin motions of a $\Gamma_8$ ground state
of Sm$^{3+}$ above $T_{NFS}\geq T_N$.\cite{Haga2004,Barla2004} The surprise is that $P_\Delta =
P_C$ i.e magnetic ordering occurs far below the pressure P$_{3+}$ where the trivalent state will be
reached at $P_{3+} \geq 10$ GPa \cite{Rohrer1982} (see figure \ref{figure8}).

An explorer at $T = 0$ K, who can look to changes in volume, conduction and magnetism, will feel
large change of volume at $P_{B-G}$ tiny change at $P_\Delta = P_C$. He will detect the concomitant
variation of the conduction and the magnetism also at $P_\Delta = P_C$. If he looks carefully in
time, he may find evidences of a sloppy matter between $P_{B-G}$ and $P_\Delta$ with unusual spin
dynamics reminiscent of the Sm$^{3+}$ configuration. At the average below $P_\Delta$ no difference
seems to occur between the $2+$ and $3+$ configurations since both present an isotropic
paramagnetism with equal sizes for their local susceptibility with either a van Vleck (2+) or a Pauli (3+)
origin.

\section{Conclusion}
The concept of Kondo impurity gives a huge impact in the physics of quantum complex systems. Its
extension  to Kondo lattice is already successful at least for the experimentalists. With slight
modifications (insertion of a pseudogap i.e. strong $n_f$ dependence of $D$), it appears possible to
track the problem of the consistency between spin and charge motion and to follow some routes in
the search for new effects and the discovery of new materials. Of course, in materials sciences,
the Kondo engineering label covers also the domain of artificial Kondo nanostructures. We can argue
of course on the validity of the engineering addenda to Kondo. At least, it reinforces that the
fundamental open questions remain. Experimentalists can built or look to new electronic and
magnetic architectures or devices. In any case, the common points are that the initial potential
has an atomic origin, the correlation length reaches nanometric size and high purity material with
an electronic mean free path above 1000 {\AA} can be achieved. Clean experiments have been already
achieved and excellent perspectives exist to go behind the standard accepted concepts to enjoy
fancy spin and charge motion and even to open new windows for applications.
It may be also worthwhile to mention that the goal to elucidate fundamental problems has boosted the development of new instrumentations which have now wide applications from astrophysics cosmology to nanophysics and material siences. The Kondo problem was a clear challenge which has required new concepts but also huge progress in the observation of low temperature physics.

JF thanks Profs. J. Friedel, T. Kasuya, K. Miyake, R. Tournier and Drs J.P. Brison, P. Haen and D. Jaccard for stimulating discussions.


\begin{thebibliography}{99} %% The number "99" means that this list has more than nine items.

\bibitem{Flouquet1978}
 J. Flouquet: \textit{Prog. Low. Temp. Physics} VII, eds. (Elsevier, North Holland, 1978).

\bibitem{Cornut1972}
 B. Cornut and B. Coqblin: Phys. Rev. B \textbf{5} (1992) 4541.

\bibitem{Flouquet1971}
J. Flouquet: Phys. Rev. Lett. \textbf{27} (1971) 515.

\bibitem{Benoit1974}
 A. Benoit, J. Flouquet, and J. Sanchez: Phys. Rev. Lett. \textbf{32} (1974) 222.

\bibitem{Hebral1977}
 B. Hebral, K. Matho, J. M. Mignot, and R. Tournier: J. Phys. Lett; \textbf{38} (1977) L347.

\bibitem{Benoit1974a}
 A. Benoit, J. Flouquet and J. Sanchez: Phys. Rev. B \textbf{1} (1974) 4213.

\bibitem{Gonzalez1973}
 F. Gonzalez Jimenez and P. Imbert: Solid State Commun. \textbf{13} (1973) 85.

\bibitem{Steglich1977}
 F. Steglich: in \textit{Festk\"{o}rperproblem} (Advances in Solid State Physics) (Vieweg, Braunschweig, 1977) ed. J. Treusch, Vol XVII, p. 319.

\bibitem{Hewson1992}
 A.C. Hewson, \textit{The Kondo problem to heavy fermions} (Cambridge University Press, 1992).

\bibitem{Doniach1977}
 S. Doniach: Physica B \textbf{91} (1977) 231.

\bibitem{Ramakrishnan1981}
 R. Ramakrishnan: \textit{Valence fluctuations in solid} (North Holland Publishing Company, 1981) eds. L.M. Falikov, W. Hanke and M.B. Maple, Vol 13.

\bibitem{Moriya1995}
 T. Moriya and Takimoto: J. Phys. Soc. Jpn. \textbf{64} (1995) 960.

\bibitem{Trees1995}
B.R. Trees, A. J. Fedro, and M. R. Norman: Phys. Rev. B \textbf{51} (1995) 6167.

\bibitem{Ideka1997}
 H. Ideka and K. Miyake: J. Phys. Soc. Jpn. \textbf{66} (1997) 3714.

\bibitem{Flouquet2004}
 J. Flouquet, to be published, Prog. Low Temp. Phys., eds A. Halperin (North Holland) (2004).

\bibitem{Burdin}
 S. Burdin, A. Georges, and D. R. Grempel: Phys. Rev. Lett. \textbf{85} (2000) 1048.

\bibitem{Nozieres1998}
  P. Nozières: Eur. J. Phys. B \textbf{6} (1998) 447 .

\bibitem{Malterre1996}
 D. Malterre, M. Grioni, and Y. Baer:  Advances in Physics \textbf{45} (1996) 299.

\bibitem{Benoit1981}
 A. Benoit, A. Berton, J. Chaussy, J. Flouquet, J. C. Lasjaunias, J. Odin, J. Pelleau, and J. Peyrard: \textit{Valence fluctuations in solid} (North Holland Publishing Company, 1981) eds. L.M. Falikov, W. Hanke and M.B. Maple, Vol 13.

\bibitem{Takke1981}
 R. Takke, M. Niksch, W. Assmus, B. L\"uthi, R. Pott, R. Schefzyk, and D. K. Wohlleben: Z. Phys. B \textbf{44} (1981) 33.

\bibitem{Jaccard1985}
 D. Jaccard and J. Flouquet: J. Magn. Magn. Mat. \textbf{47-48} (1985) 45.

\bibitem{Zhu2003}
 L. Zhu, M. Garst, A. Rosch, and Q. Si: Phys. Rev. Lett. \textbf{91} (2003) 066404.

\bibitem{Flouquet2002}
 J. Flouquet, P. Haen, S. Raymond, D. Aoki, and G. Knebel: Physica B \textbf{319} (2002) 251.

\bibitem{Fisher1991}
 R.A. Fisher, C. Marcenat, N. E. Phillips, P Haen, F. Lappiere, P. Lejay, J. Flouquet and J. Voiron: J. Low Temp. Phys. \textbf{84} (1991) 49.

\bibitem{Lacerda1989}
 A. Lacerda, A. de Visser, L. Puech, P. Haen, and J. Flouquet: Phys. Rev. B \textbf{40} (1989) 11429.

\bibitem{Aoki1995}
 H. Aoki, M. Takashita, S. Uji, T. Terashima, K. Maezawa, R. Settai, Y. Onuki: Physica B \textbf{206-207} (1995) 26.

\bibitem{Julian1994}
 S.R. Julian, F. S. Tautz, G. J. McMullan, and G. G. Lonzarich: Physica B \textbf{199-200} (1994) 63.

\bibitem{Payer1993}
 K. Payer P Haen, J. M. Laurant, J. M. Mignot, and J. Flouquet: Physica B \textbf{186-188} (1993) 503.

\bibitem{Haen1996}
 P. Haen \textit{et al.}: J. Phys. Soc. Jpn. \textbf{65} (1996) suppl. B, 27 .

\bibitem{Raymond1999}
 S. Raymond, D. Raoelison, S. Kambe, L. P. Regnault, B. Fak, R. Calemzuk, J. Flouquet, P. Haen, and P. Lejay: Physica B \textbf{259-261} (1999) 48.

\bibitem{Sato2004}
 M. Sato \textit{et al.}: to be published.

\bibitem{Custers2003}
 C.V. Custers, P. Gegenwart, H. Wilhelm, K. Neumaier, Y. Tokiwa, O. Trovarelli, C. Geibel, F. Steglich, C. Pepin, and P. Coleman: Nature (London) \textbf{424} (2003) 524.

\bibitem{Satoh2001}
 H. Satoh and F.J. Ohkawa: Phys. Rev. B \textbf{63} (2001) 184401.

\bibitem{Amato1993}
 A. Amato, B. Baines, R. Feyerherm, J. Flouquet, F. N. Gygax, P. Lejay, A. Schenk, and U. Zimmermenn: Physica B \textbf{186-188} (1993) 276.

\bibitem{Amato1989}
 A. Amato, D. Jaccard, J. Sierro, P. Haen, P. Lejay, and J. FLouquet: J. Low Temp. Phys. \textbf{77} (1989) 195.

\bibitem{Knafo}
 W. Knafo, PhD thesis, Univerity Grenoble (2004)

\bibitem{Knebel2002}
 G. Knebel, D. Braithwaite, P. C. Canfield, G. Lapertot, and J. Flouquet: Phys. Rev. B \textbf{65} (2002) 024425.

\bibitem{Demuer2002}
 A. Demuer, A. T. Holmes, and D. Jaccard: J. Phys. Condens. Matter \textbf{14} (2002) L529.

\bibitem{Araki2002}
 S. Araki, Nakashima, R. Settai, T. C. Kobayashi, and Y. Onuki: J. Phys. Condes. Matter \textbf{14} (2002) 377.

\bibitem{Glazkov2004}
 V. Glazkov and G. Knebel, private communication.

\bibitem{Holmes2004}
 A.T. Holmes, D. Jaccard, and K. Miyake: Phys. Rev. B \textbf{69} (2004) 024508.

\bibitem{Jaccard1992}
 D. Jaccard, K. Behnia, and J. Sierro: Phys. Lett. A \textbf{163} (1992) 475.

\bibitem{Knebel2004}
 G. Knebel et al, to be published (2004).

\bibitem{Thompson2001}
 J.D. Thompson, R. Movshovich, Z. Fisk, F. Bouquet, N. J. Curro, R. A. Fisher, P. C. Hammel, H. Hegger, M. F. Hundley, M. Jaime, P. G. Pagliuso, C. Petrovic, N. E. Phillips, and J. L. Sarrao: J. Magn. Magn. Mat. \textbf{226-230} (2001) 5.

\bibitem{Llobet2004}
 A. Llobet, J. S. Gardner, E. G. Moshopoulou, J. M. Mignot, M. Nicklas, W. Bao, N. O. Moreno, I. N. Gonscharenko, J. L. Sarrao, and J. D. Thompson: Phys. Rev. B \textbf{69} (2004) 024403.

\bibitem{Barla2004}
 A. Barla, J. P. Sanchez, Y. Haga, G. Lapertot, B. P. Doyle, O. Leupold, R. R\"uffer, M. M. Abd-Elmeguid, R. Lengsdorf, and J. Flouquet: Phys. Rev. Lett. \textbf{92} (2004) 066401.

\bibitem{Haga2004}
 Y. Haga, A. Barla, J. Derr, B. Salce, G. Lapertot, I. Sheikin, K. Matsubayashi, N. K. Sato, and J. Flouquet, to be published (2004).

\bibitem{Barla}
 A. Barla, J. P . Sanchez, J. Derr, B. Salce, G. Lapertot, J. Flouquet, B. P. Doyle, O. Leupold, R. R\"uffer, M. M. Abd-Elmeguid, and R. Lengsdorf, to be published in J. Phys. Condens. Matter, cond-mat/0406166

\bibitem{Rohrer1982}
J. R\"{o}hrer et al, Valence instabilities, eds. P. Wachter and H. Boppart (Amsterdam, North
Holland) p. 215 (1982).

\bibitem{Dallera2004}
 C. Dallera, E. Annexe, J. P. Rueff, M. Grioni, G. Yanko, L. Braicovich, A. Barla, J. P. Sanchez, R. Gusmeroli, A. Polenzana, L. Degiorgi, and G. Lapertot: to be published in J. Condens. Matter (20)


\end{thebibliography}
\end{document}